\documentclass[aps,prapplied,epsfigure,twocolumn,superscriptaddress]{revtex4-2}
\usepackage[colorlinks=true,linkcolor=blue,urlcolor=blue,citecolor=blue,pdfusetitle]{hyperref}
\usepackage[utf8]{inputenc}
\usepackage[english]{babel}
\usepackage{amsmath}
\usepackage[caption = false]{subfig}
\usepackage{graphicx,epstopdf}
\usepackage{blindtext}
\usepackage[table,xcdraw]{xcolor}
\usepackage{lipsum}
\usepackage{amsfonts}
\usepackage{bbm}
\usepackage{amssymb}
\usepackage{enumerate}
\usepackage{color}
\usepackage{latexsym}
\usepackage{physics}
\usepackage{times,txfonts}
\usepackage{ulem}
\usepackage{amsmath}
\usepackage{braket}

\newcommand{\Mcal}{\mathcal{M}}

\newcommand{\1}{\mathbbm{1}}

\definecolor{MyPurple}{RGB}{204, 0, 102}

\newcommand{\interpro}[2]{\langle #1 | #2 \rangle}

\newcommand{\SubFig}[2]{\ref{#1}{\color{blue}#2}}

\usepackage{color} 
\usepackage{orcidlink}

\begin{document}
	
	\title{Native conditional $i$SWAP operation with superconducting artificial atoms}
	
	\newcommand{\siqse}{Shenzhen Institute for Quantum Science and Engineering, Southern University of Science and Technology, Shenzhen, Guangdong 518055, China}
	\newcommand{\physustech}{Department of Physics, Southern University of Science and Technology, Shenzhen, Guangdong 518055, China}
	\newcommand{\gdpkl}{Guangdong Provincial Key Laboratory of Quantum Science and Engineering, Southern University of Science and Technology, Shenzhen, Guangdong 518055, China}
	\newcommand{\szkl}{International Quantum Academy, Futian District, Shenzhen, Guangdong 518048, China}
	
	\newcommand{\UFSCar}{Departamento de F\'{i}sica, Universidade Federal de S\~ao Carlos, Rodovia Washington Lu\'{i}s, km 235 - SP-310, 13565-905 S\~ao Carlos, SP, Brazil}	
	\newcommand{\SU}{Department of Physics, Stockholm University, AlbaNova University Center 106 91 Stockholm, Sweden}
	\newcommand{\Nice}{Universit\'e C\^ote d'Azur, CNRS, Institut de Physique de Nice, 06560 Valbonne, France}
	
	\author{Chang-Kang Hu}
	\email{huck@sustech.edu.cn}
	\affiliation{\siqse}\affiliation{\szkl}\affiliation{\gdpkl}
	
	\author{Jiahao Yuan}
	\affiliation{\siqse}\affiliation{\szkl}\affiliation{\gdpkl}\affiliation{\physustech}
	
	\author{Bruno A. Veloso}
	\affiliation{\UFSCar}
	
	\author{Jiawei Qiu}
	\affiliation{\siqse}\affiliation{\szkl}\affiliation{\gdpkl}\affiliation{\physustech}
	\author{Yuxuan Zhou}
	\affiliation{\siqse}\affiliation{\szkl}\affiliation{\gdpkl}\affiliation{\physustech}
	\author{Libo Zhang}
	\affiliation{\siqse}\affiliation{\szkl}\affiliation{\gdpkl}
	\author{\\Ji Chu}
	\affiliation{\siqse}\affiliation{\szkl}\affiliation{\gdpkl}
	\author{Orkesh Nurbolat}
	\affiliation{\siqse}\affiliation{\szkl}\affiliation{\gdpkl}
	
	\author{Ling Hu}
	\affiliation{\siqse}\affiliation{\szkl}\affiliation{\gdpkl}
	\author{Jian Li}
	\affiliation{\siqse}\affiliation{\szkl}\affiliation{\gdpkl}
	\author{Yuan Xu}
	\affiliation{\siqse}\affiliation{\szkl}\affiliation{\gdpkl}
	\author{Youpeng Zhong}
	\affiliation{\siqse}\affiliation{\szkl}\affiliation{\gdpkl}
	\author{Song Liu}
	\affiliation{\siqse}\affiliation{\szkl}\affiliation{\gdpkl}
	\author{\\Fei Yan}
	\email{yanfei@baqis.ac.cn}
	\affiliation{\siqse}\affiliation{\szkl}\affiliation{\gdpkl}
	\affiliation{Present address: Beijing Academy of Quantum Information Sciences, Beijing 100193, China}
	\author{Dian Tan}
	\email{tand@sustech.edu.cn}
	\affiliation{\siqse}\affiliation{\szkl}\affiliation{\gdpkl}
	\author{R. Bachelard~\orcidlink{0000-0002-6026-509X}} 
	\affiliation{\Nice}
	\affiliation{\UFSCar}
	\author{Alan C. Santos~\orcidlink{0000-0002-6989-7958}}
	\email{ac\_santos@df.ufscar.br}
	\affiliation{\UFSCar}
	\affiliation{\SU}
	\author{C. J. Villas-Boas~\orcidlink{0000-0001-5622-786X}} 
	\affiliation{\UFSCar}
	
	\author{Dapeng Yu}
	\affiliation{\siqse}\affiliation{\szkl}\affiliation{\gdpkl}\affiliation{\physustech}
	
	\begin{abstract}
	Controlling the flow of quantum information is a fundamental task for quantum computers, which is unfeasible to realize on classical devices. Coherent devices which can process quantum states are thus required to route the quantum states that encode information. 
	In this paper we demonstrate experimentally the smallest quantum transistor with a superconducting quantum processor which is composed of a \textit{collector} qubit, an \textit{emitter} qubit, and a coupler (transistor gate). The interaction strength between the collector and emitter qubits is controlled by the frequency and state of the coupler, effectively implementing a quantum switch. Through the coupler-state-dependent Heisenberg (inherent) interaction between the qubits, a single-step (native) conditional $i$SWAP operation can be applied. To this end,
	we find that it is important to take into consideration higher energy level for achieving a native and high-fidelity transistor operation. By reconstructing the Quantum Process
		Tomography, we obtain an operation fidelity of $92.36\%$ when the transistor gate is open ($i$SWAP implementation) and $95.23 \%$ in the case of closed gate (identity gate implementation). The architecture has strong potential in quantum information processing applications with superconducting qubits.
	\end{abstract}

	\maketitle

	
	\section{Introduction} 
	
	Thanks to the advent of semiconductors physics and advances in solid state physics~\cite{Ridley:Book,Phillips:Book}, devices based on quantum effects have been used to design first-generation quantum technologies. These devices exploit tunneling and band-structure of complex systems and have been used to build transistors with nanometer physical gate~\cite{Desai:16}, although the information flow obeys classical laws. Furthermore, spin-based transistors have allowed for the manipulation and engineering of atom-like spins at an elementary level, with applicability to quantum information science~\cite{Igor:04}. Since transistors are fundamental hardware components in classical computers, similar devices able to process bits of quantum information have been proposed as quantum analogs of such devices~\cite{Marchukov:16,sun2018,KasperPRA:22,shan2018,Kasper:22}, including a universal model for adiabatic quantum computation~\cite{Williamson:15}. In this sense, quantum transistors may be useful components to control the flow and processing of information in operations of quantum computation~\cite{Bacon:17}. Such conditional state transfer has been experimentally demonstrated in liner optics system~\cite{patel2016quantum} and superconducting circuits system~\cite{gao2019entanglement}, but the processes are not deterministic or native respectively. These are two important requirements for quantum efficient transistor.
	
	Among the physical platforms candidate to the construction of quantum processors, superconducting circuit systems have stood out as a cutting-edge technology for the realization of a scalable quantum computer~\cite{arute2019quantum, gong2021quantum, wu2021strong}, in addition to being a promising platform to efficiently study quantum many-body physics~\cite{You:11,Wang:20,ZhangKe:21,Zanner:22}. Although superconducting circuits present a multilevel structure (artificial atoms), they can also be operated as two-level systems (qubits) to implement tunable interaction between the parts of a superconducting processor~\cite{Li:20,Han:20,Feng:20,collodo2020implementation, Qiu:21, stehlik2021tunable, Yan:18, sung2021realization}. In this scenario, \textit{frequency-tunable couplers} are one of the most popular architectures for superconducting quantum computing~\cite{Yan:18,sung2021realization,Xu:20,Rasmussen:21}, where the gate control is done by applying an external flux through the coupler. However, these kind of frequency tunbale control interactions is sensitive to flux noise~\cite{Braumuller:20,Campbell:22}, such that alternatives are required for robust operations. In this paper, we exploit the idea of a quantum transistor, as proposed by D. Bacon \textit{et al}~\cite{Bacon:17}, and a coupler-state-dependent interaction to implement conditional operations.

\begin{figure}[t!]
	\includegraphics[width=\linewidth]{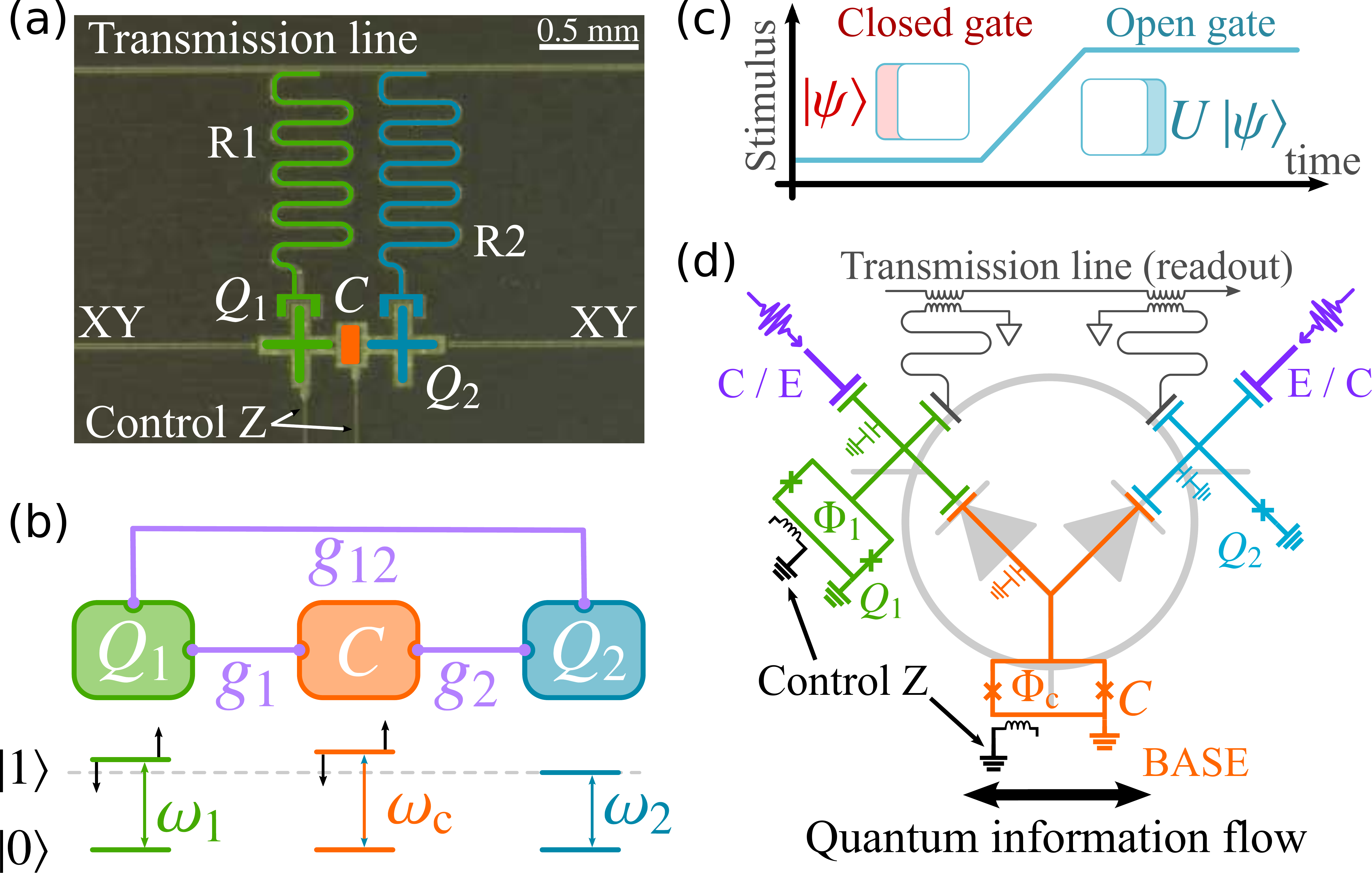}
	\caption{(a) Chip with the superconducting quantum circuits used in our experiment. (b) Schematic diagram of the three interacting components of the circuit, with the tunable control qubit in-between the two target qubits. (c) Generic representation of a conditional quantum operation $U$ implemented by a quantum transistor-like mechanism. (d) Schematic representation of the superconducting quantum transistor, where the information is encoded in the collector qubit and sent to the emitter qubit in a coupler-state-dependent way. The reversibility of the quantum evolution allows to achieve a two-way transistor for quantum information.}\label{Fig1}
\end{figure}
	
	By using two superconducting qubits allied with a frequency-tunable coupler~\cite{Qiu:21} (see Figs.~\SubFig{Fig1}{a} and~\SubFig{Fig1}{b}), we demonstrate a coherent conditional implementation of an $i$SWAP gate. 
	The analysis of the conditional operation is done through the effective dynamics of the superconducting qubits with tunable interactions, which allows to identify the critical role of the third energy level for the dynamics. As a first result, although the two-level approach is quite precise in the description of the system effective dynamics~\cite{Yan:18, Li:20,Han:20,Feng:20,collodo2020implementation, Qiu:21, stehlik2021tunable, sung2021realization}, our experimental data supports that under
	special conditions the three-level approach needs to be considered to control precisely the real two-qubit dynamics.
	As an application and second result, we exploit a state-switchable transfer of information in superconducting processors, by effectively realizing for the first time a quantum Bacon-Flammia-Crosswite transistor-like operation~\cite{Bacon:17}. This specific transistor 
	implements a quantum operation $U$ according to a transistor gate state, that can be changed through an external stimulus, as schematized in Fig.~\SubFig{Fig1}{c}). Our device, in Fig.~\SubFig{Fig1}{d}, differs from previous proposals in literature since the control gate is encoded into multi-qubits state~\cite{Marchukov:16,Loft:18,Bacon:17}, so that our study constitutes the first experimental realization of the single-qubit quantum switch, that is, the smallest quantum transistor proposed so far in superconducting integrated circuits.

	\section{Modelling and two-qubit effective dynamics} 
	
	Our superconducting circuit is schematically presented in Figs.~\ref{Fig1}{\color{blue}a} and~\ref{Fig1}{\color{blue}b}, where the tunable coupler qubit ($C$) lies in-between two Xmon superconducting artificial atoms ($Q_1$ and $Q_2$). Using an external flux (chain control $Z$) across the SQUID loop, the frequencies of the atom $Q_1$ and of the coupler are made tunable, whereas the one of the atom $Q_2$ remains fixed.

	\begin{table}[b]
		\centering  
		\scriptsize  
		\caption{Parameters of the superconducting qubits used in our experiment.}  
		\begin{ruledtabular}
			\begin{tabular}{rccccccp{0cm}}
				\hspace{0.5cm} 
				{\bf \small Qubit}  & 
				{\small $\mathrm{\bf Freq}^{\mathrm{\bf Max}}$}  &  
				{\small $\mathrm{\bf Freq}^{\mathrm{ \bf Idling}}$} & 
				{\bf \small $\alpha$} & 
				{ \small $T_{1}^{\mathrm{\bf Idling}}$} & 
				{\bf \small $T_{2}^{\mathrm{ \bf Idling}}$ }& 
				{\bf \small $T_{\mathrm{\bf 2,Echo}}^{\mathrm{\bf Idling}}$ }&  \\ 
				\hline 
				\hline\\[-2mm]  
				
				Qubit 1 &  5.230 GHz  &  4.670 GHz  & -222 MHz&  6.51 us&  0.54 us & 3.83us &\\
				Coupler &  8.831 GHz  &  6.183 GHz  & -378 MHz&  4.06 us&  0.27 us & 2.42us &\\
				Qubit 2 &     ---     &  4.619 GHz  & -242 MHz&  6.58 us&  7.43 us & 13.02us &\\
				
			\end{tabular}
		\end{ruledtabular}
		\label{tab_para}
	\end{table}
	
	%
	
	As a preliminary step to realize the transistor, let us discuss the nature of our artificial atoms:	they cannot be treated as simple two-level systems, as it was done previously~\cite{Yan:18, Li:20,Han:20,Feng:20,collodo2020implementation, Qiu:21, stehlik2021tunable, sung2021realization}. The system is thus described by a Hamiltonian of the form $H\!=\!H_{0} + V_{0}$~\cite{Xu:20}, where
	\begin{align}
	\label{H0}
	H_{0} &= \hbar\sum\nolimits_{i=1,2,c} \left(\omega_i \, a_i^\dagger a_i + \frac{\alpha_i}{2} \, a_i^\dagger a_i^\dagger a_i a_i \right) ,
	\end{align}
	is the bare Hamiltonian of the atoms and the coupler, with $a_i^\dagger$ and $a_i$ their creation and annihilation operators, and $\omega_{i}$ the transition frequency between the ground state and the first excited state. $\alpha_{i}$ is the energy level anharmonicity, with $\alpha_{i}$ big enough corresponding to the two-level system, and $\alpha_{i}=0$ to the harmonic oscillator. The term
	\begin{align}
	V_{0} = \hbar\sum\nolimits_{i = 1,2}\left[g_{i} \left( a_i^\dagger a_{\mathrm{c}} + a_i a_{\mathrm{c}}^\dagger \right)\right] + \hbar g_{12}\left( a_1^\dagger a_2 + a_1 a_2^\dagger \right) ,
	\end{align}
	describes the interaction between the components of the superconducting circuit, with $g_{i}$ the (capacitive) coupling strength between the $i$th atom and the coupler. Such design allows for an atom-coupler coupling $g_{i}$ much stronger than the capacitive atom-atom coupling $g_{12}$ ($g_{12}\!\ll\!g_{i}$). More precisely, we measure coupling strengths of order $g_{1}\approx110$~MHz, $g_{2}\approx105$~MHz and $g_{12}\approx7.5$~MHz. The other parameters of the circuit are shown in Table~\ref{tab_para}.
	
	As a first result, we show that assuming the superconducting artificial atoms behave as two-level systems leads to inaccurate predictions for the coupled dynamics, which can in turn affect the operation of the circuit. In the explored regime, each atom must be considered at least as a three-level system, where the anharmonicity of the third energy level plays an important role when the state of the coupler changes. Let us set the two atoms $Q_1$ and $Q_2$ at the same frequency ($\omega_{1}\!=\!\omega_{2}\!=\!\omega$), and tune the coupler frequency toward the dispersive regime $|\Delta|\!\gg\!g_{i}$~\cite{Yan:18,Li:20,Han:20,Feng:20}, with $\Delta = \omega_{1}-\omega_{\mathrm{c}}$. The effective Hamiltonian then reads $\tilde{H}_{\text{eff}}^{\ket{n}_{\mathrm{c}}} = \hbar \tilde{g}_{\text{eff}}^{\ket{n}_\mathrm{c}}(\Delta)[\sigma_{1}^{-}\sigma_{2}^{+} + \sigma_{2}^{-}\sigma_{1}^{+}]$, where the effective coupling coefficients $\tilde{g}_{\text{eff}}^{\ket{n}_\mathrm{c}}(\Delta)$ read:
	\begin{align}
	\tilde{g}_{\text{eff}}^{\ket{n}_\mathrm{c}}(\Delta) = g_{12} + g_{1}g_{2}\left(\frac{2}{\Delta-\delta_{n1}\alpha_{\mathrm{c}}} - \frac{1}{\Delta}\right), \label{Eq-EffectiveH}
	\end{align}
	with $\delta_{nm}$ the Kronecker delta symbol. This result is consistent with the two-level approach adopted in Refs.~\cite{Yan:18,Li:20,Han:20,Feng:20} {\it when the coupler is in the ground state}. However, when it is in the first excited state, the anharmonicity $\alpha_{\mathrm{c}}$ comes into play, and  that it affects the effective coupling: This state-dependent coupling has a direct impact on the operation of our device, as we shall now see.
	
	\begin{figure}[t!]
		\includegraphics[width=\linewidth]{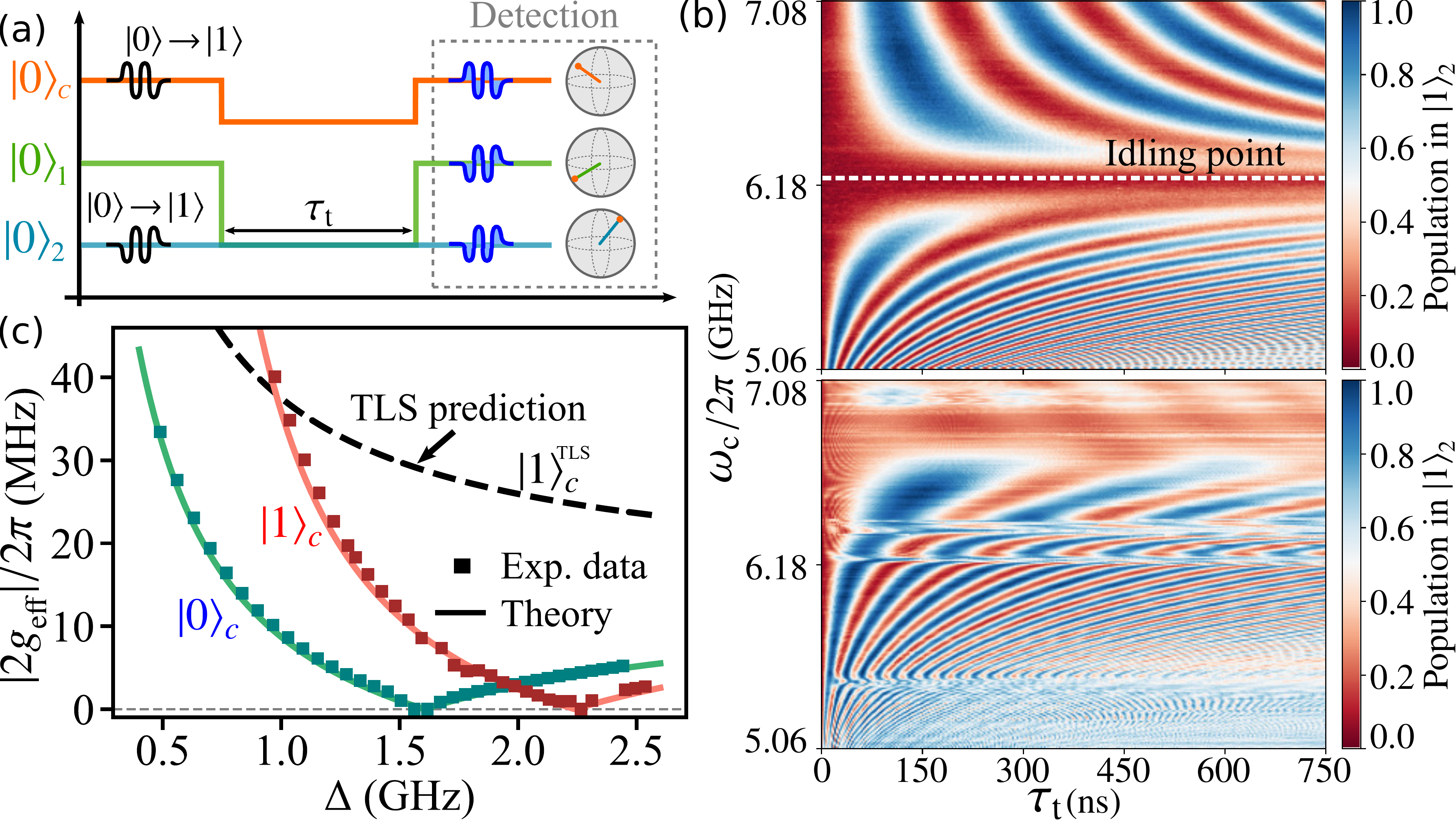}
		\caption{Effective coupling strength between two target qubits under the control of the quantum state and frequency of the coupler. (a) Pulse sequence of the experimental implementation to measure the effective coupling strength. (b) Experimental data for resonant exchange between $\ket{0}_{\mathrm{c}}\ket{01}_{12}$ and $\ket{0}_{\mathrm{c}}\ket{10}_{12}$ (top), and for $\ket{1}_{\mathrm{c}}\ket{01}_{12}$ and $\ket{1}_{\mathrm{c}}\ket{10}_{12}$ (bottom), where the y-axis is the coupler frequency. By fitting the vacuum Rabi oscillation, we extract the effective coupling strength $|2\tilde{g}_{\text{eff}}^{\ket{n}_\mathrm{c}}(\Delta) |/2\pi$, whose behavior as function of $\Delta$ is shown in (c). Symbols denote the experimental data, while curves represent theoretical prediction. Full lines describe the expectation from Eq.~(3) and the dotted line corresponds to the prediction of the two-level approach.}\label{Fig2}
	\end{figure}
	
	\begin{figure*}[t!]
		\includegraphics[width=\linewidth]{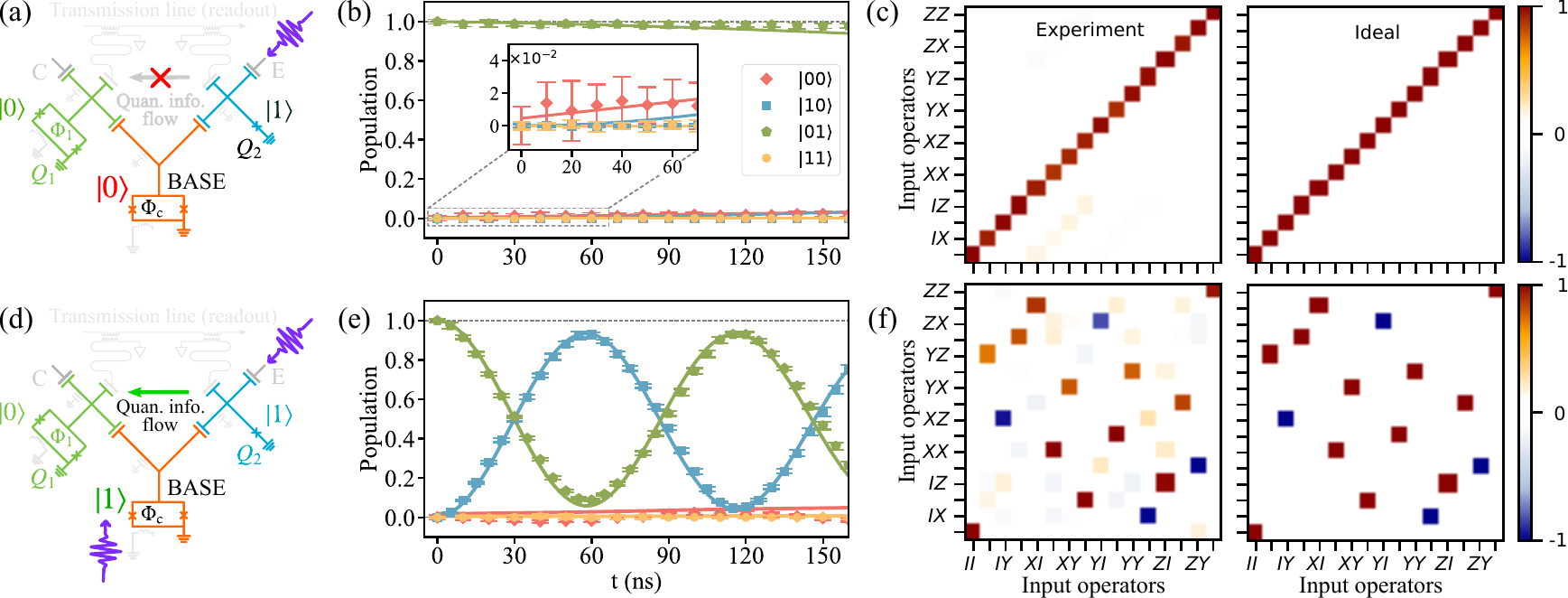}
		\caption{(a,d) Sketch of the quantum transistor operation, which allows for the state transfer from $Q_{2}$ to $Q_{1}$ (and vice-versa) when an external stimulating field promotes an excitation (opening the gate) and de-excitation (closing the gate) in the coupler. (b,e) Theoretical (curves) and experimental (dots) data for transfer and blockade performance of the transistor. The inset in figure (b) shows the low population in the state $\ket{1 0}_{12}$, which constitutes the final state in the transfer process. The slow increase of the population in the state $\ket{0 0}_{12}$ stems from the relaxation of $Q_{2}$. (c,f) Experimental and ideal, respectively, quantum process tomography matrix of the transistor operation.}
		\label{Fig3}
	\end{figure*}
	
	The contribution of the third level is probed in our setup by implementing the procedure sketched in Fig.~\ref{Fig2}{\color{blue}a}. We use a power splitter to combine a DC signal and a pulse signal to control the frequency of the frequency-tunable atoms and coupler. The DC signal is a biased signal used to set the idling frequency of $Q_1$ and of the coupler, which remain unchanged during the experiment. We then set the coupler frequency idling point to completely turn off the effective coupling between $Q_1$ and $Q_2$ (white horizontal dotted line in Fig.~\ref{Fig2}{\color{blue}b}) when the coupler is in the ground state, and we set the $Q_1$ frequency idling point to be about $50$~MHz above that of $Q_2$. In this case, the interaction between $Q_1$, $Q_2$ and the coupler can be neglected, and the computational basis approximates the eigenstates of the bare system. Through this procedure, one can then use the single-qubit gate ($\pi$-pulse) to efficiently prepare the system in the state $\ket{0}_{\mathrm{c}}\ket{01}_{12}$ or $\ket{1}_{\mathrm{c}}\ket{01}_{12}$.
	The effective interaction is then turned on by putting the atom $Q_1$ at resonance with $Q_2$. After the interaction step, we switch back the frequency of the atoms and coupler to their initial configuration (idling point) to measure the fidelity in the binary state detection (see Ref.~\cite{SupInf} for further details). 
	
	The above sequence is used to monitor the oscillation of population between $Q_1$ and $Q_2$: this only requires to measure the population of the two-qubit state $\ket{0 1}_{12}$, independently on the coupler state, which we present in Fig.~\ref{Fig2}{\color{blue}b}. The coupling coefficient $\tilde{g}_{12}^{\ket{n}_{\mathrm{c}}}$ is extracted from these oscillations by fitting the oscillating dynamics of this population with a sine function. Its dependence on the detuning $\Delta$ and the coupler state $\ket{n}_{\mathrm{c}}$ is shown in Fig.~\ref{Fig2}{\color{blue}c}. As mentioned before, for the coupler in the ground state, the measured behaviour of the coupling $\tilde{g}_{12}^{\ket{0}_{\mathrm{c}}}$ is in very good agreement with the theoretical prediction even if the system is considered to possess only two levels, since, as there is only one excitation in the whole system, the higher levels do not play a role. On the other hand, the two-level model fails to predict the behavior of an excited coupler, $\tilde{g}_{12}^{\ket{1}_{\mathrm{c}}}$, which may lead to a loss in high control of two-qubit operations. Differently, our three-level approach leads to an effective coupling for which theory (see Eq.~\eqref{Eq-EffectiveH}) and experiment (see Fig.~\ref{Fig2}{\color{blue}c}) agree very well. This leads to a dramatic increase in the controllability on the system evolution and, consequently, in the robustness and fidelity of the derived quantum computation processes.
	
	Although we can turn on and turn off the dynamics in our system using different quantum states of the coupler, it is worth mentioning that we can turn-off the dynamics even when the coupler is in the excited state. As shown by the red curve in Fig.~\ref{Fig2}{\color{blue}c}, even when the coupler is in the excited state, a new ``off-state" of the coupler (identity gate) can be obtained by adequately choosing the coupler frequency to a new idle point such that $\Delta^{\prime} \approx 2.25~$GHz.
	
	
	\section{Conditional $i$SWAP operation}
	
	Let us now demonstrate how a quantum transistor can be achieved using our circuit, which illustrates the coherent manipulation of the quantum information flow through the system. To this end, we exploit the dependence of the coupling on the coupler state, which allows us to realize a gate. The qubits $Q_1$ and $Q_2$ now correspond to the collector and emitter, in which we encode and read-out the quantum information, while the coupler acts as the control gate. First we prepare the information to be transferred in the emitter $Q_2$, while the coupler (transistor base) remains in its ground state, see Fig.~\ref{Fig3}{\color{blue}a}. We set the coupler frequency at $\omega_{\mathrm{c}}\!=\!6.183 $~GHz and we turn on the effective interaction between $Q_1$ and $Q_2$. In Fig.~\ref{Fig3}{\color{blue}b}, we present the population of the different states when the information encoded corresponds to a single excitation $\ket{\psi}_{2}\!=\!\ket{1}_{2}$. The population then remains blockaded in state $\ket{01}_{12}$, that is, the information is not transferred to the collector. 
	
	The fidelity of the operations implemented in our system is computed by the ideal and experimental Quantum Process Tomography (QPT). To this end, we prepare 16 two-qubit product states $\ket{\phi_1}\ket{\phi_2}$, where the state of each qubit $\ket{\phi_n}$ is selected from the set \{$\ket{0}$, $\ket{1}$, $\ket{0}+\ket{1}$, $\ket{0}+i\ket{1}$\}. The density matrix of the output state is determined through the standard two-qubit state tomography, by measuring each qubit along the $\sigma_x$, $\sigma_y$, and $\sigma_z$ axes of the Bloch sphere~\cite{Nielsen:Book}. Following the correction of measurement errors, we reconstruct the 16x16 experimental process matrix. To calculate the fidelity between the experimental and ideal quantum processes we employ the following formula $F = \mathrm{Tr}(\chi_\mathrm{exp} \chi_\mathrm{ideal})$, where $\chi_\mathrm{exp}$ and $\chi_\mathrm{ideal}$ correspond to the experimental and ideal process matrices, respectively.
	
	Through reconstruction of the QPT matrix for this operation~(Fig.~\ref{Fig3}{\color{blue}c}), one reaches fidelity around $F_{\mathrm{clos}}\!=\!95.23(52)\%$ for the closed gate.
	In the case of the open gate (see Fig.~\ref{Fig3}{\color{blue}d}), the coupler is initially excited and a coherent effective coupling between $Q_1$ and $Q_2$ ($2\tilde{g}_{12}/2\pi\!=\!8.45$~MHz) drives the system through a Heisenberg excitation exchange. This dynamics can be observed through the coherent flow of excitation from $Q_2$ to $Q_1$, as illustrated in Fig.~\ref{Fig3}{\color{blue}c}. This shows that our proposal constitutes a two-way transistor, where the information can coherently flow both ways throughout the system, thus being a signature of the quantumness of our transistor~\cite{SupInf}. From a fundamental point of view, this symmetric behavior is a direct consequence of the unitary nature of the dynamics and of the quantum version of the recurrence theorem of Poincaré~\cite{Bocchieri:57}. From the corresponding ideal and experimental QPT for the open gate, the fidelity for this operation around $F_{\mathrm{open}}\!=\!92.36(75)\%$ is obtained (see Ref.~\cite{SupInf} for further details). The duration of the operation shown in Fig. 3e is approximately $59$~ns. The efficiency of this dynamics is here limited by the damping timescale of the emitter, as shown in Table~\ref{tab_para}.
	
	As one of the main contributions of this work, we show how the control discussed above allows one to implement $i$SWAP operations between the two qubits, a crucial feature for achieving universal quantum computation through Heisenberg exchange interactions~\cite{DiVincenzo:00Nature,Jeremy:02}. Instead of monitoring the coherent exchange of population between $Q_1$ and $Q_2$, the Fig.~\ref{Fig3}{\color{blue}f} shows the  quantum process tomography for the qubits when the coupler starts in the excited state. The right panel in Fig.~\ref{Fig3}{\color{blue}f} shows the ideal  quantum process tomography expected result for an ideal $i$SWAP operation between $Q_1$ and $Q_2$, while the left panel of Fig.~\ref{Fig3}{\color{blue}f} corresponds to the experimental data.
	
	It shows that, thanks to the particular feature of our circuit our device (see Fig.~\SubFig{Fig1}{d}) provides us two important features. First, conditional coherent control of information which depends on the quantum state of a \textit{single} coupler (working as a \textit{quantum switch}), instead of using multi-qubit states to perform that task~\cite{Marchukov:16,Loft:18,Bacon:17}. Second, the realization of a one-step two-qubit conditional $i$SWAP gate, using the inherent Heisenberg interactions of the system, constitutes the first experimental implementation of such a gate in a \textit{native} way. It is an important step for quantum computation, which is one of the challenging, but promising, technologies in contemporary science~\cite{fedorov2012implementation, reed2012realization, patel2016quantum, gao2019entanglement}.

	\section{Conclusion}
	
	In conclusion, we have presented a quantum transistor based on a three-component superconducting circuit, where high-fidelity operations are made possible by accounting for the multi-level structure of the superconducting artificial atoms. The anharmonicity between the two first excited states of these atoms plays an important role, which manifests in the dependence of the dynamics on the coupler state. In particular, a two-level description is valid only when the coupler remains in its ground state, so this anharmonicity will inevitably affect the operation of the transistor. Our implementation of the transistor-like behavior with Xmon artificial atoms confirm the role of this anharmonicity, with theoretical and experimental results in very good agreement. Our results thus allow us to design a precisely-controllable mechanism for the conditional $i$SWAP operation between two parts of a superconducting circuit. Also, as sketched in Fig.~\ref{Fig1}{\color{blue}d}, the two atoms act as collector and emitter, while the coupler controls the information flow (i.e., the control gate), so the coherent control of quantum states transfer is achieved between the two qubits. Thus, in our system the coupler is a part of the quantum circuit that acts as a quantum switch to control information flow and two-qubit $i$SWAP gates.

	
	While frequency division multiplexing is an effective technology to circumvent the challenge of controlling qubits by using their distinct frequencies, enabling a single control line to manage multiple qubits, flux adjustment and frequency division multiplexing methods are incompatible~\footnote{Such a incompatibility comes from the fact that a flux signal is a low frequency signal and do not have carrier wave. So, we can not use different carrier frequencies to address the qubit/coupler.}. Differently, the proposed strategy of tuning the coupler state are compatible with frequency division multiplexing. It enables the control of the coupler state via frequency division multiplexing to achieve desired operations without adjusting the coupler frequency. This eliminates the need for a separate flux control line in sample design, conserving resources and offering a viable solution for large-scale, scalable quantum computing.
	
	\textit{Note added.--} While completing this project, we became aware of a complementary demonstration of how the control of the coupler state can be useful to implement a CZ gate on fluxonium qubits~\cite{Simakov:23}.
	

	\begin{acknowledgments}
		This work was supported by the Key-Area Research and Development Program of Guang-Dong Province (Grant No. 2018B030326001), the National Natural Science Foundation of China (12205137, 12004167, 11934010, 1801661), the China Postdoctoral Science Foundation (Grant No. 2020M671861, {2021T140648}), the Guangdong Innovative and Entrepreneurial Research Team Program (2016ZT06D348), the Guangdong Provincial Key Laboratory (Grant No.2019B121203002), the Natural Science Foundation of Guangdong Province (2017B030308003), and the Science, Technology and Innovation Commission of Shenzhen Municipality (JCYJ20170412152620376, KYTDPT20181011104202253), and the NSF of Beijing (Grants No. Z190012). A.C.S., B. A. V., C.J.V.-B., and R.B. acknowledge the financial support of the São Paulo Research Foundation (FAPESP) (Grants No. 2018/15554-5, No. 2019/22685-1, No. 2019/11999-5, No. 2019/13143-0 and No. 2021/10224-0) and the Coordenação de Aperfeiçoamento de Pessoal de Nível Superior (CAPES/STINT), Grants No.
		88881.304807/2018-01, and No. 88887.512104/2020-00. R.B. and C.J.V.-B. benefitted from the support of the National Council for Scientific and Technological Development (CNPq) Grants No. 302981/2017-9, No.		409946/2018-4, and No. 311612/2021-0. C.J.V.-B. is also thankful for the support from the Brazilian National Institute 		of Science and Technology for Quantum Information (INCTIQ/CNPq) Grant No. 465469/2014-0.
		
		C.-K.H., J.Y. and B.A.V. contributed equally to this work.
	\end{acknowledgments}
    \vspace{-0.3cm}
	\appendix
    \section{Experimental setup}

\begin{figure*}[t!]
		\centering
		\includegraphics[scale=0.82]{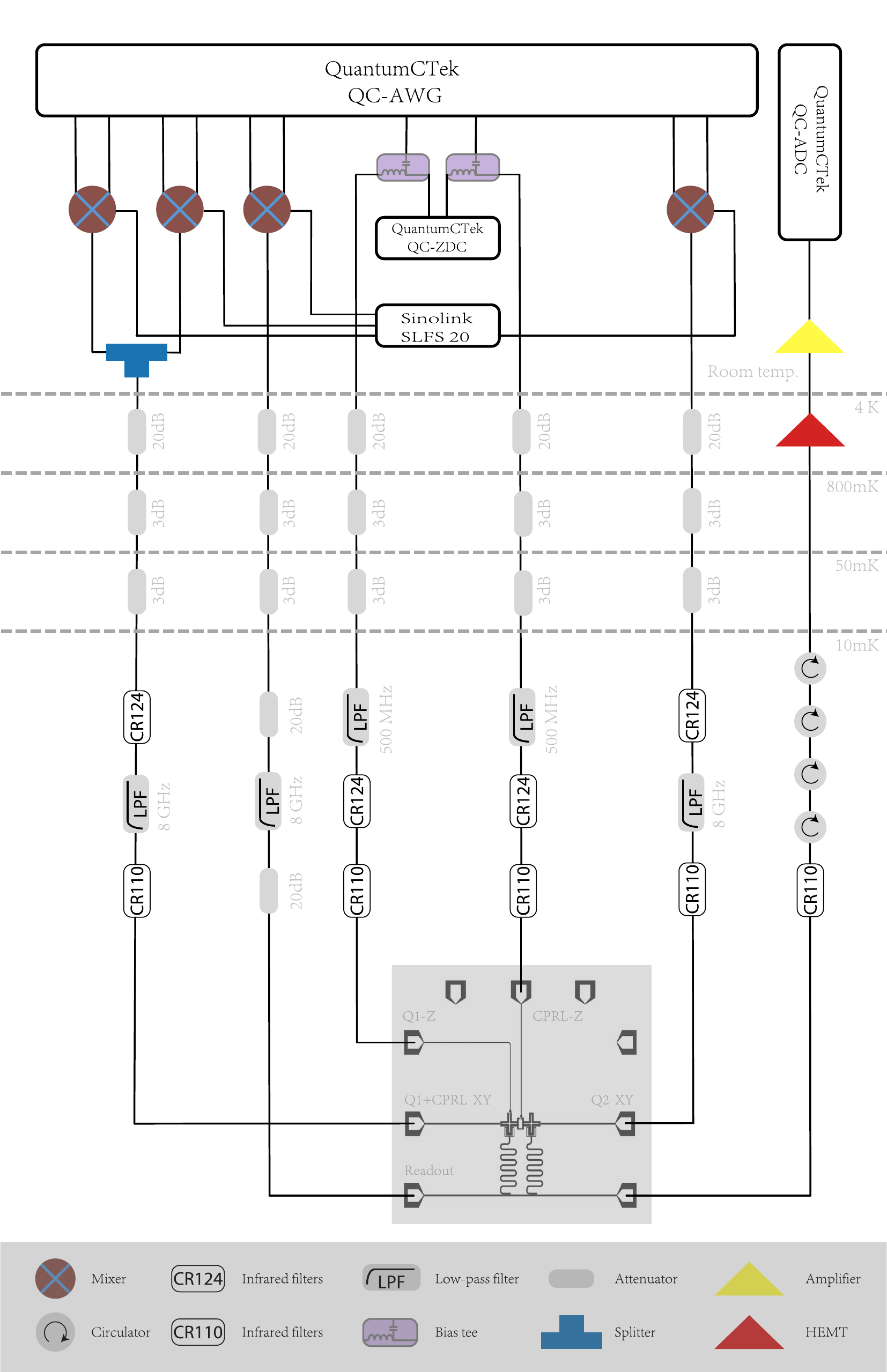}
		\caption{Electronics and sample schematic diagram of the experimental setup}
		\label{Figure_Setup-SM}
	\end{figure*}

	The transmon superconducting chip is installed inside a BlueFors XLD-1000 dilution refrigerator system, and its base temperature is under 10 mK. We magnetically shield the chip with a Cryoperm cylinder.
	The Electronics and sample diagram are shown in Fig.~\ref{Figure_Setup-SM}{\color{blue}}.
	To perform the standard circuit-QED measurements, we apply microwave to the input port of the readout transmission line. Four microwave isolators are placed before the high electron-mobility transistor (HEMT) amplifier to prevent noise from higher-temperature stages. After being amplified by the HEMT amplifier at the 4K stage and a low noise amplifier at room temperature, the readout signal will be downconverted, and the demodulated IQ signals will be digitized by analog-to-digital converters.
	All of the control electronics which is used to apply the XY and Z controls of the qubits and coupler are at room temperature. The control signals and the readout signals are programmed in Labber software and sent to the QuantumCTek arbitrary waveform generator (AWG). Then, the corresponding microwave pulses generated by the AWGs will be mixed with different local oscillators (LO), respectively. Here, the LO is supported by a commercial Multichannel coherence microwave generator Sinolink SLFS20. In the experiment, we need to add extra DC Z control signals to set the idling points of the tunable qubit and the coupler.
	So, we use a bias-tee to combine the DC signals and the fast Z control pulse signals.

	\section{Readout errors correction}
\begin{figure}[t!]
		\centering
		\includegraphics[width=.8\linewidth]{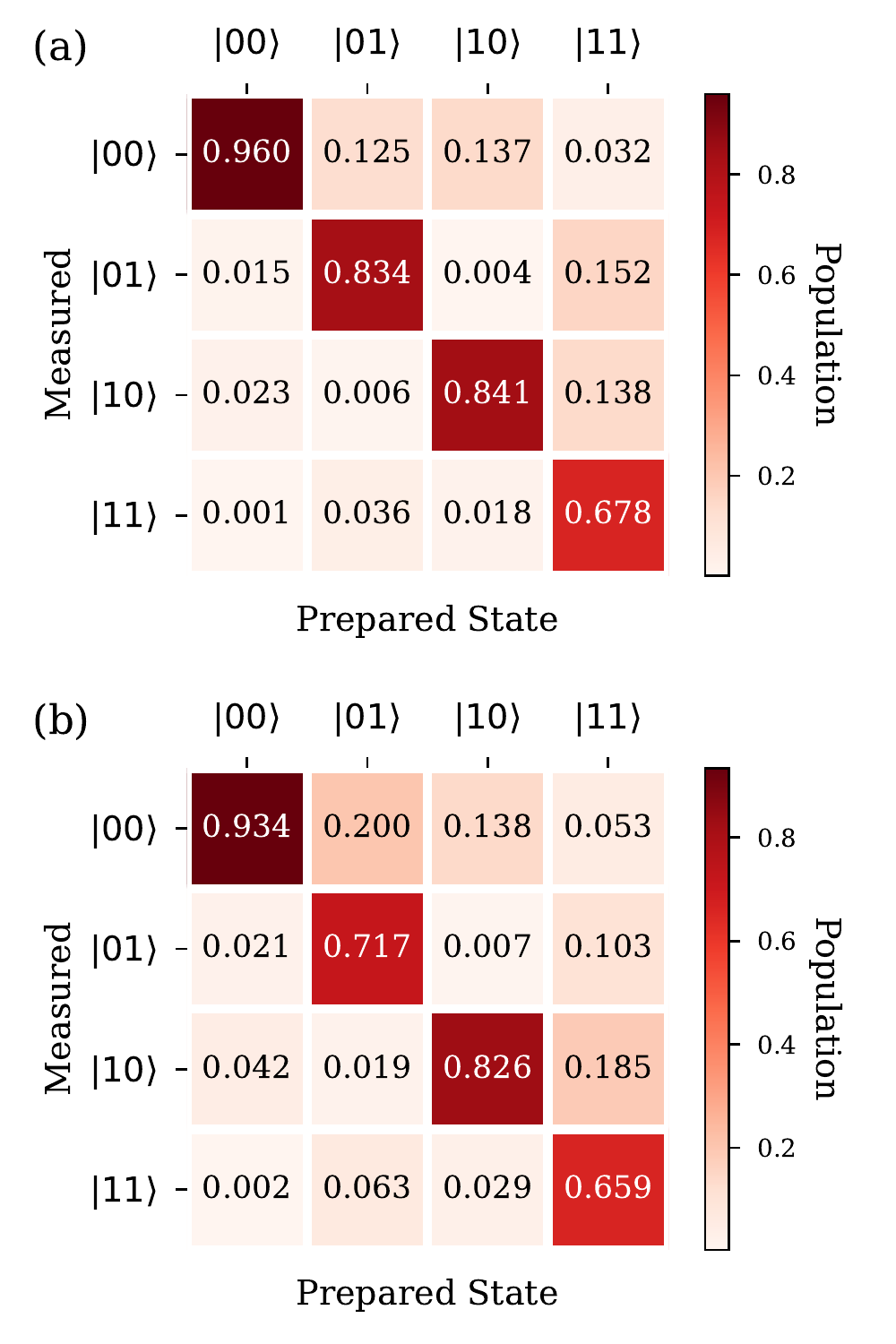}
		\caption{Readout errors characterization.  Each column in the graph is the measured population distribution of the prepared initial state.  The transfer matrix $\Mcal$ is simply the population matrix.  (a) Measured when coupler is in state $|0\rangle$.  (b) Measured when coupler is in state $|1\rangle$.}
		\label{fig:m_table_off}
	\end{figure}

	Before characterizing the performance of our quantum transistor, we need to identify and correct the readout errors of the system.  Readout errors consist of the incorrect mapping error of single-qubit and the cross-talk error between qubits~\cite{Dewes2012, Ficheux2021}.  Here, a transfer matrix $\Mcal$ is adapted to correct both readout errors simultaneously.  And the transfer matrix working on the joint readout population is written as
	\begin{equation}
		\begin{pmatrix}
			p_{00}'\\
			p_{10}'\\
			p_{01}'\\
			p_{11}'
		\end{pmatrix}
		=
		\begin{pmatrix}
			M_{00}&M_{01}&M_{02}&M_{03} \\
			M_{10}&M_{11}&M_{12}&M_{13} \\
			M_{20}&M_{21}&M_{22}&M_{33} \\
			M_{30}&M_{31}&M_{32}&M_{33} \\
		\end{pmatrix}
		\begin{pmatrix}
			p_{00}\\
			p_{10}\\
			p_{01}\\
			p_{11}
		\end{pmatrix},
		\label{eq:readout_error_correction}
	\end{equation}
	where $p_{ij}'$ are the measurement qubit populations, and $p_{ij}$ are the corrected qubit populations. To find $\Mcal$, we prepare two qubits in states $\ket{00}$, $\ket{01}$, $\ket{10}$ and $\ket{11}$ successively, and perform single-shot measurements in $\ket{00}$, $\ket{01}$, $\ket{10}$ and $\ket{11}$ bases.  As shown in Fig.~\ref{fig:m_table_off}, we get $\Mcal$ directly from the measured population matrix.  Then, we are able to correct measured population with equation $\vec{p} = \Mcal^{-1} \vec{p}'$.

	\section{Two-level system effective Hamiltonian}

	As discussed in the main text, assuming a two-level system (TLS) to deduce the effective Hamiltonian is not adequate to describe the effective dynamics of the system. To verify this point, we first define the raising and lowering operators for each of our  atom $i$ as $\sigma^{+}_{i} = \ket{1}_i\bra{0}$ and $\sigma^{-}_{i}=\ket{0}_i\bra{1}$, respectively, as well as the population operators $ \sigma^{0}_i = \ket{0}_i\bra{0} $ and $ \sigma^{1}_i = \ket{1}_i\bra{1} $. Then, assuming $ \omega_{1} = \omega_{2} = \omega $, we can write the Hamiltonian presented in Eqs. (1) and (2) of the main text as
	\begin{align}\label{EQ:TLS_Hamiltonian}
		H &= \sum_{i = 1,2} \hbar\omega \sigma_{i}^{1} +\hbar g_{ic}(\sigma_{i}^{+}\sigma_{c}^{-}+h.c.)\nonumber \\ &+ \hbar g_{12}(\sigma_{1}^{+}\sigma_{2}^{-}+h.c.)+\hbar\omega_{c}\sigma_{c}^{1}.
	\end{align}

	Then, using a unitary transformation $H_{I}=UHU^{\dagger}-H_{0},$ with $H_{0}=\hbar\omega(\sigma_{1}^{1}+\sigma_{2}^{1})+\hbar\omega_{c}\sigma_{c}^{1}$ and $U(t)=e^{-iH_{0}t/\hbar}$, casting Eq.~(\ref{EQ:TLS_Hamiltonian}) to the interaction picture we find
	\begin{align}\label{EQ:TLS_InteractionHamiltonian}
		H_{\mathrm{I}}&=\hbar g_{1}(\sigma_{1}^{+}\sigma_{c}^{-}e^{i\Delta t} + \mathrm{h.c.})+
		\hbar g_{2}(\sigma_{2}^{+}\sigma_{c}^{-}e^{i\Delta t} + \mathrm{h.c.})\nonumber\\ &+ \hbar g_{12}(\sigma_{1}^{-}\sigma_{2}^{+}+\sigma_{2}^{-}\sigma_{1}^{+}),
	\end{align}
	where we have defined $\Delta = \omega - \omega_{c} $. In the interaction picture, we can apply the Rotating Wave Approximation (RWA) to the Hamiltonian, 
	which here translates as
	\begin{align}\label{EQ:EffectiveHamiltonian}
		H_{\mathrm{eff}} &  \approx \frac{1}{\hbar}\left[-iH_{I}(t)\int_{0}^{t}H_{I}(t')dt'\right]_{RWA},
	\end{align}
	where $ i $ stands for the imaginary unity, and where we have assumed that $ |\Delta|\gg g_{k} $ to neglect the fast-oscillating terms. That leads to the TLS effective Hamiltonian
	\begin{align}\label{TLS_EffectiveHamiltonian}
		H_{\mathrm{eff}}&=\hbar\frac{g_{1}g_{2}}{\Delta}\left[ h_0 + (\sigma_{1}^{-}\sigma_{2}^{+}+\sigma_{2}^{-}\sigma_{1}^{+})(\sigma_{c}^{0}-\sigma_{c}^{1})\right]\nonumber\\
		&+\hbar g_{12}(\sigma_{1}^{-}\sigma_{2}^{+}+\sigma_{2}^{-}\sigma_{1}^{+}),
	\end{align}
	where $h_0\!=\!(\sigma_{1}^{1}+\sigma_{2}^{1})\sigma_{c}^{0}-(\sigma_{1}^{0}+\sigma_{2}^{0})\sigma_{c}^{1}$, with $\sigma_{k}^{n}\!=\!\ket{n}\bra{n}_{k}$, is an energy shift term which does not promote any population transference in the system.
	When we consider only the interaction terms of Eq. (\ref{TLS_EffectiveHamiltonian}), we can identify the effective coupling
	\begin{align}\label{TLS_EffectiveCoupling}
		g_{\mathrm{eff}}^{\ket{n}_{\mathrm{c}}}(\Delta) &= g_{12} + (-1)^n \frac{g_{1}g_{2}}{\Delta},
	\end{align}
	where the $(-1)^n$ term is a signature  that the effective coupling depends on the coupler state.

	\section{Three-level system effective Hamiltonian}
	To evaluate the effects of the third level of energy on the dynamics, we need to consider the anharmonic terms in Eq. (1) of the main text. Let us proceed similarly to the TLS case: First we define $ \tilde{H}_0 =  \sum_{j=1,2,c} \left[\omega_j \, a_j^\dagger a_j + \frac{\alpha_j}{2} \, a_j^\dagger a_j^\dagger a_j a_j\right]$ as the unperturbed Hamiltonian and use the unitary operator $ U(t)=e^{\frac{-i\tilde{H}_{0}t}{\hbar}} $ to write the Hamiltonian in the interaction picture. However, in this case, we define the operators $ \Sigma_{j} $ for each artificial atom, and proceed with the transformations
	\begin{align}
		a_{j} &\rightarrow \Sigma_{j}^{-} = \sum_{k=1}^{2} \sqrt{k}\ket{k-1}\bra{k} ,\nonumber\\ a_{j}^{\dagger} &\rightarrow \Sigma_{j}^{+} = \sum_{k=1}^{2} \sqrt{k}\ket{k}\bra{k-1}.
	\end{align}
	This allow us to write the Hamiltonian in the interaction picture
	\begin{align}\label{EQ:3lvl_InteratonPicture_1}
		\tilde{H}_{I} &= \sum_{k = 1,2}\hbar g_{i} \left( U^{\dagger}(t)\Sigma^{+}_k U(t)U^{\dagger}(t)\Sigma^{-}_{\text{c}} U(t) + h.c. \right) \nonumber \\ &+ \hbar g_{12}\left( U^{\dagger}(t)\Sigma^{+}_1 U(t)U^{\dagger}(t) \Sigma^{-}_2U(t) + h.c. \right).
	\end{align}

	Defining the operator $P_{nm}^{(j)}\!=\! \ket{n} \bra{m}_j$ and manipulating the Eq.~\eqref{EQ:3lvl_InteratonPicture_1}, the Hamiltonian assumes the form
	\begin{align}
		\tilde{H}_{I} &= H_{1,c}(t) + H_{2,c}(t) + H_{2}(t)
	\end{align}
	with
	\begin{align}
		H_{k,c}(t) &= \hbar g_{k}\left[e^{i (\omega_{k} - \omega_{\text{c}})t} P_{10}^{(k)} P_{01}^{(\text{c})} + e^{i (\omega_{k}-\tilde{\omega}_{\text{c}})t} \sqrt{2}P_{10}^{(k)} P_{12}^{(\text{c})} \right.\nonumber \\ &\left. +
		e^{i (\tilde{\omega}_{k}-\omega_{\text{c}})t} \sqrt{2} P_{21}^{(k)}  P_{01}^{(\text{c})} + 2e^{i (\tilde{\omega}_{k}-\tilde{\omega}_{\text{c}})t} P_{21}^{(k)} P_{12}^{(\text{c})}  + h.c. \right]
		\nonumber \\
		H_{2}(t) &= \hbar g_{12} \left[ e^{i (\omega_{1} - \omega_{2})t} P_{10}^{(1)} P_{01}^{(2)} + e^{i (\omega_{1}-\tilde{\omega}_{2})t} \sqrt{2}P_{10}^{(1)} P_{12}^{(2)} \right.\nonumber \\ &\left.+ e^{i (\tilde{\omega}_{1}-\omega_{2})t} \sqrt{2} P_{21}^{(1)}  P_{01}^{(2)} + 2 e^{i (\tilde{\omega}_{1}-\tilde{\omega}_{2})t} P_{21}^{(1)}  P_{12}^{(2)} + h.c. \right]
	\end{align}
	where, $ \tilde{\omega}_i = \omega_i + \alpha_i $. Assuming, for simplicity, that $ \omega_{1}=\omega_{2}=\omega $, and $ \alpha_{1} = \alpha_{2} = \alpha $ we can write the Hamiltonian as
	\begin{align}\label{3lvl_InteractionHamiltonian}
		\tilde{H}_{I} & =H_{0}+\sqrt{2}\hbar g_{12}\left[e^{it\alpha}\left(P_{21}^{(1)}P_{01}^{(2)}+P_{01}^{(1)}P_{21}^{(2)}\right) \right.\nonumber \\
		&\left.\hspace{2.2cm}+e^{-it\alpha}\left(P_{12}^{(1)}P_{10}^{(2)}+P_{10}^{(1)}P_{12}^{(2)}\right)\right]\nonumber \\
		& +\sum_{k=1,2}\hbar g_{k}\left[e^{i\Delta t}P_{10}^{(k)}P_{01}^{(\text{c})}+e^{i\tilde{\Delta}t}\sqrt{2}P_{10}^{(k)}P_{12}^{(\text{c})} + \right.\nonumber \\&\left.\hspace{1.4cm}+e^{i\text{\ensuremath{\tilde{\Delta}'}}t}\sqrt{2}P_{21}^{(k)}P_{01}^{(\text{c})}+2e^{i\tilde{\tilde{\Delta}}t}P_{21}^{(k)}P_{12}^{(\text{c})}+h.c.\right] ,
	\end{align}
	where $ H_{0}=g_{12}\left(P_{10}^{(1)}P_{01}^{(2)}+2P_{21}^{(1)}P_{12}^{(2)}+h.c.\right)$ is the time-independent part of the Hamiltonian, and $\tilde{\Delta}=\omega-\tilde{\omega}_{c}$,
	$\tilde{\Delta}'=\tilde{\omega}-\omega_{c}$ and $\tilde{\tilde{\Delta}}=\tilde{\omega}-\tilde{\omega}_{c}$ are the detunnings. In the interaction picture, obtaining the effective Hamiltonian is similar to the TLS case, where we use Eq.~(\ref{EQ:EffectiveHamiltonian}) and assume that $\Delta\gg g_{1},g_{2}$ and $\alpha,\alpha_{c} \gg g_{12}$ to be able to neglect the fast oscillating terms. The final result is
	\begin{widetext}
		\begin{align}\label{EQ:3lvl_EffectiveHamiltonian}
			H_{\mathrm{eff}}&= \sum_{k,m=1,2} \hbar g_{k}g_{m}\left[\frac{1}{\omega-\omega_{c}}(P_{10}^{(k)}P_{01}^{(m)}P_{00}^{(c)}-P_{01}^{(k)}P_{10}^{(m)}P_{11}^{(c)})\right.+\frac{2}{\omega-\tilde{\omega}_{c}}(P_{10}^{(k)}P_{01}^{(m)}P_{11}^{(c)}-P_{01}^{(k)}P_{10}^{(m)}P_{22}^{(c)})\nonumber \\
			+ & \left.\frac{2}{\tilde{\omega}-\omega_{c}}(P_{21}^{(k)}P_{12}^{(m)}P_{00}^{(\text{c})}-P_{12}^{(k)}P_{21}^{(m)}P_{11}^{(\text{c})})+\frac{4}{\tilde{\omega}-\tilde{\omega}_{c}}(P_{21}^{(k)}P_{12}^{(m)}P_{11}^{(\text{c})}-P_{12}^{(k)}P_{21}^{(m)}P_{22}^{(\text{c})})\right]\nonumber \\
			+ & \hbar\frac{2g_{12}^{2}}{\alpha}\left(-P_{11}^{(1)}P_{11}^{(2)}+P_{22}^{(1)}P_{00}^{(2)}+P_{20}^{(1)}P_{02}^{(2)}+h.c.\right)+ \hbar g_{12}\left(P_{10}^{(1)}P_{01}^{(2)}+2P_{21}^{(1)}P_{12}^{(2)}+h.c.\right).
		\end{align}
	\end{widetext}
	The difference between Eq.~(\ref{EQ:3lvl_EffectiveHamiltonian}) and Eq.~(\ref{TLS_EffectiveHamiltonian}) is quite clear: When neglecting the effects of the third level of energy, a significant part of the effective dynamics is lost, which leads to a difference in the expression for the effective coupling, when the coupler is in the first excited state. To verify that, we calculate the effective Hamiltonians taking in account only the interaction terms, and find
	\begin{subequations}
		\begin{align}
			\tilde{H}_{\mathrm{eff}}^{|0\rangle_{c}}&=\hbar\left(\frac{g_{1}g_{2}}{\Delta}+g_{12}\right)(P_{10}^{(1)}P_{01}^{(2)}+h.c.)\nonumber \\ &+2\hbar\left(g_{12}+\frac{g_{1}g_{2}}{\Delta+\alpha}\right)(P_{21}^{(1)}P_{12}^{(2)}+h.c.) \label{EQ:3lvl_EffectiveHamiltonian_0_c} \\
			\tilde{H}_{\mathrm{eff}}^{|1\rangle_{c}}&=
			\hbar\left[2g_{12}-2g_{1}g_{2}\left(\frac{1}{\Delta+\alpha}-\frac{2}{\alpha-\alpha_{c}}\right)\right] (P_{12}^{(1)}P_{21}^{(2)}+h.c.)
			\nonumber \\
			&+   \hbar\left[g_{12}-g_{1}g_{2}\left(\frac{1}{\Delta}-\frac{2}{\Delta-\alpha_{c}}\right)\right](P_{01}^{(1)}P_{10}^{(2)}+h.c.)  \label{EQ:3lvl_EffectiveHamiltonian_1_c}.
		\end{align}
	\end{subequations}

	Given that no transition occurs to the second excited state, only the $ (P_{01}^{(1)}P_{10}^{(2)}+h.c.) $ terms will contribute to the effective coupling, which leads to the effective two-level system Hamitonian
	\begin{align}
		\tilde{H}_{\mathrm{eff}}^{\ket{n}_{\mathrm{c}}}(\Delta)=\hbar \tilde{g}_{\mathrm{eff}}^{\ket{n}_{\mathrm{c}}}(\Delta)(\sigma_{1}^{-}\sigma_{2}^{+}+\sigma_{2}^{-}\sigma_{1}^{+}),
	\end{align}
	where
	\begin{align}\label{EQ:3lvl_EffectiveCoupling}
		\tilde{g}_{\mathrm{eff}}^{\ket{n}_{\mathrm{c}}}(\Delta) = g_{12} + g_1 g_2\left(
		\frac{2}{\Delta - \delta_{n1}\alpha_{c}}- \frac{1}{\Delta}\right),
	\end{align}
	with $ \delta_{n1} $ the Kronecker delta. Comparing the expressions for the effective couplings obtained in Eq.(\ref{TLS_EffectiveCoupling}) for the TLS and now in Eq.~(\ref{EQ:3lvl_EffectiveCoupling}), we can verify that when the coupler is in the ground state, both expressions are equivalent, however, when we have an excitation on the coupler, the TLS expression differs from the three-level approach by a factor $ (2g_1g_2)/(\Delta - \alpha_{c}) $. This term can only be negligible in the limit of a genuine two-level system, that is, when we have anharmonicity large enough ($ \alpha_{c} \rightarrow \infty $).

	\subsection{Effective dynamics and Quantumness of the device}

	In this section we discuss the theory used to find the value of $\tilde{g}_{\mathrm{eff}}^{\ket{1}_{\mathrm{c}}}(\Delta)$, as observed from the experimental data. The effective Hamiltonian in the three-level system Eq.~(\ref{EQ:3lvl_EffectiveHamiltonian_1_c}) can be used to verify the inversion of excitation between $ Q_1 $ and $ Q_2 $ when the system is set on the initial state $ \ket{\Psi(0)}\!=\!\ket{1}_{\mathrm{c}} $ of the coupler. In this case, the Hamiltonian can be written in the $ \{\ket{10}, \ket{01}\} $ basis as
	\begin{align}
		\tilde{H}_{\mathrm{eff}} = \tilde{g}_{\mathrm{eff}}^{\ket{1}_{\mathrm{c}}}(\Delta)\left[\begin{array}{cc}
			0 & 1\\
			1 & 0
		\end{array}\right],
	\end{align}
	which has the eigenenergies $E_{\pm}\!=\!\pm \tilde{g}_{\mathrm{eff}}^{\ket{1}_{\mathrm{c}}}(\Delta)$, associated with eigenstates $\ket{\Psi_{\pm}}\!=\!(\ket{01}_{12}\pm\ket{10}_{12})/\sqrt{2}$. By computing the evolved state $ \ket{\Psi(t)} = e^{-iHt/\hbar}\ket{\Psi(0)} $, from the initial state $ \ket{\Psi(0)}\!=\!\ket{1}_{\mathrm{c}} \ket{10}_{12}  $, we find
	\begin{align}
		\ket{\Psi(t)}=\cos(\tilde{g}_{\mathrm{eff}}^{\ket{1}_{\mathrm{c}}}t)~\ket{01}_{12}-\text{i}\sin(\tilde{g}_{\mathrm{eff}}^{\ket{1}_{\mathrm{c}}}t)~\ket{10}_{12} .
	\end{align}

	\begin{figure}[t!]
		\centering
		\includegraphics[width=\linewidth]{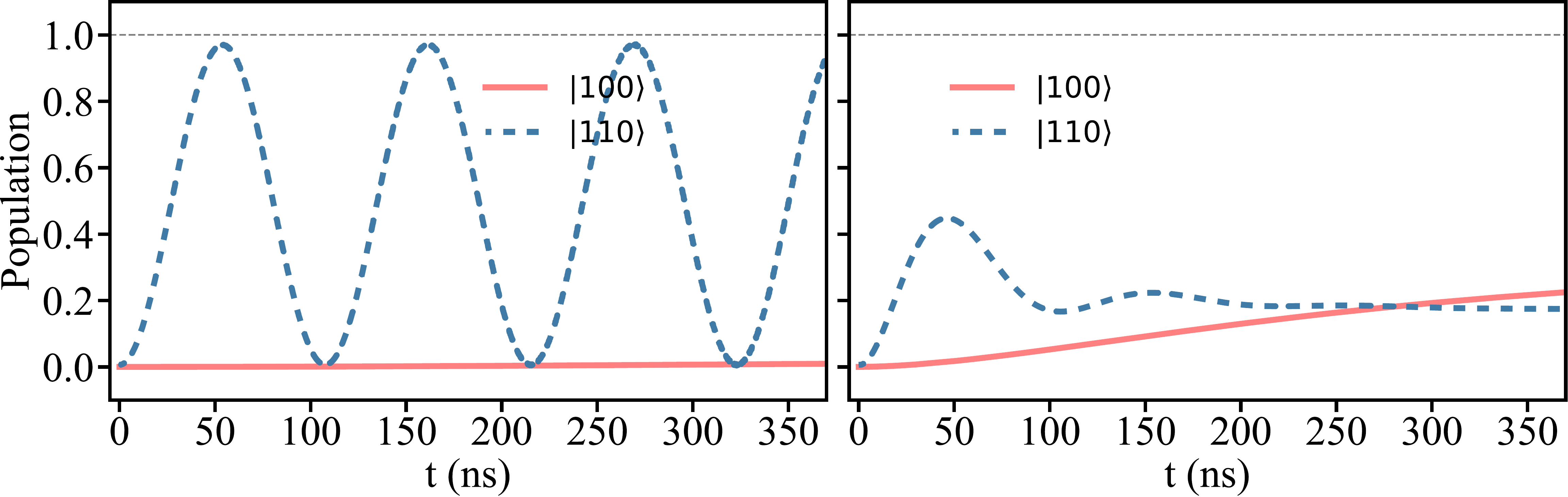}
		\caption{Fidelity of transfer and blockade when (a) no decoherence acts on the system, and (b) the coupler is affected by decoherence that brings the system into the classical realm. To this simulation we use the same parameters as in the main text with $\Gamma\!=\!g_{1}$.}
		\label{fig:classical}
	\end{figure}

	Therefore, by measuring the probability of getting the system in states $\ket{10}$ and $\ket{01}$, we find
	\begin{align}
		P_{1}(t) &= |\interpro{\Psi(t)}{10}|^2 = \cos^2(\tilde{g}_{\mathrm{eff}}^{\ket{1}_{\mathrm{c}}}t) , \\
		P_{2}(t) &= |\interpro{\Psi(t)}{01}|^2 = \sin^2(\tilde{g}_{\mathrm{eff}}^{\ket{1}_{\mathrm{c}}}t) .
	\end{align}
	Thus, from the experimental data for $P_{1}(t)$ and $P_{2}(t)$, we can measure the value of $\tilde{g}_{\mathrm{eff}}^{\ket{1}_{\mathrm{c}}}$ by fitting the curves $P_{1}(t)$ and $P_{2}(t)$.

	We now briefly explore the quantumness of the device proposed here, where we investigate the performance of the system under strong decoherence, which brings the system into a classical realm. We focus on the simpler case where we assume that the noise affects only the coupler. In this case the system is governed by the master equation
	\begin{align}
		\dot{\rho}(t) = -\frac{i}{\hbar} [H,\rho(t)] + \frac{\Gamma}{2} \left[ 2a^{\dagger}_{\mathrm{c}}a_{\mathrm{c}} \rho(t)a_{\mathrm{c}}a^{\dagger}_{\mathrm{c}} - \{a_{\mathrm{c}}a^{\dagger}_{\mathrm{c}}a^{\dagger}_{\mathrm{c}}a_{\mathrm{c}},\rho(t)\} \right] ,
	\end{align}
	where the last operator describes the dephasing effect on the system with rate $\Gamma$. When $\Gamma$ is strong enough the coherent transport of information is drastically deteriorated as shown in Fig.~\ref{fig:classical}. This shows that, due to the loss of coherence, the device cannot explore quantum effects and the performance becomes worse than the case where no decoherence acts on the system.


%
	
\end{document}